\documentclass[pdflatex,sn-apa]{sn-jnl}% APA Reference Style 
\usepackage{graphicx}%
\usepackage{multirow}%
\usepackage{amsmath,amssymb,amsfonts}%
\usepackage{amsthm}%
\usepackage{mathrsfs}%
\usepackage[title]{appendix}%
\usepackage{xcolor}%
\usepackage{textcomp}%
\usepackage{manyfoot}%
\usepackage{booktabs}%
\usepackage{algorithm}%
\usepackage{algorithmicx}%
\usepackage{algpseudocode}%
\usepackage{listings}%
\usepackage[shortlabels]{enumitem}
\usepackage{xspace}
\usepackage{newfloat}
\usepackage{listings}
\floatstyle{ruled}
\newfloat{listing}{tb}{lst}{}
\floatname{listing}{Listing}
\usepackage{rotating}
\usepackage{mdframed}
\usepackage{soul}
\usepackage[colorinlistoftodos,prependcaption,textsize=tiny]{todonotes}
\begin{document}

%% Manually defined commands:
\newcommand{\yusuf}[1]{\textcolor{orange}{#1}}
\newcommand{\lx}[1]{\textcolor{red}{#1}}
\newcommand{\lxnote}[1]{\textcolor{red}{TODO: #1}}
\newcommand{\approach}{Transducer Tuning\xspace}
\newcommand{\transducer}{Transducer\xspace}

\title[Article Title]{Transducer Tuning: Efficient Model Adaptation for Software Tasks Using Code Property Graphs}


\author{\fnm{Imam Nur Bani} \sur{Yusuf}}\email{imamy.2020@phdcs.smu.edu.sg}

\author{\fnm{Lingxiao} \sur{Jiang}}\email{lxjiang.smu.edu.sg}

\affil{
\orgdiv{School of Computing and Information Systems},
\orgname{Singapore Management University}
% \orgaddress{\street{Street}, 
% \city{City}, 
% \postcode{100190}, 
% \state{State}, 
% \country{Country}
}


\abstract{
Large language models have demonstrated promising performance across various software engineering tasks. While fine-tuning is a common practice to adapt these models for downstream tasks, it becomes challenging in resource-constrained environments due to increased memory requirements from growing trainable parameters in increasingly large language models.
We introduce \approach, a technique to adapt large models for downstream code tasks using Code Property Graphs (CPGs). Our approach introduces a modular component called \transducer that enriches code embeddings with structural and dependency information from CPGs. The Transducer comprises two key components: Graph Vectorization Engine (GVE) and Attention-Based Fusion Layer (ABFL). GVE extracts CPGs from input source code and transforms them into graph feature vectors. ABFL then fuses those graph feature vectors with initial code embeddings from a large language model.
% to enrich it with structural and dependency information from CPGs. 
By optimizing these transducers for different downstream tasks, our approach enhances the models without the need to fine-tune them for specific tasks.
We have evaluated \approach on three downstream tasks: code summarization, assert generation, and code translation. Our results demonstrate competitive performance compared to full parameter fine-tuning while reducing up to 99\% trainable parameters to save memory.
%\todo{can we make it clearer how much memory is saved?}.
\approach also remains competitive against other fine-tuning approaches (e.g., LoRA, Prompt-Tuning, Prefix-Tuning) while using only 1.5\%-80\% of their trainable parameters. Our findings show that integrating structural and dependency information through Transducer Tuning enables more efficient model adaptation, making it easier for users to adapt large models in resource-constrained settings.
}




\maketitle

\section{Introduction}\label{sec1}

Large language models have demonstrated promising performance across various software engineering tasks, such as code generation \citep{DBLP:conf/acl/ZanCZLWGWL23}, code summarization \citep{DBLP:journals/corr/abs-2311-10372, DBLP:journals/corr/abs-2401-00288}, and code repair \citep{DBLP:conf/icse/XiaWZ23, DBLP:conf/sigsoft/JinSTSLSS23}. "Pretrain, then fine-tune" is a common practice to adapt the models for downstream tasks \citep{DBLP:conf/naacl/DevlinCLT19, DBLP:conf/emnlp/FengGTDFGS0LJZ20, DBLP:conf/iclr/GuoRLFT0ZDSFTDC21, DBLP:conf/naacl/AhmadCRC21, DBLP:conf/emnlp/WangLGB0H23}.
These models are often pretrained by large organizations on huge corpora, learning code syntax, semantics, and patterns \citep{DBLP:conf/issta/ShiW0DH0023, DBLP:journals/corr/abs-2403-14608}.
Users can then adapt the pretrained models to their specific needs through fine-tuning on domain-specific data.
Model fine-tuning often works by iteratively updating the models' parameters through gradient descent \citep{rumelhart1986learning}, using input-output examples for specific downstream tasks. Through this optimization process, the model's parameters gradually align with the patterns and objectives of the downstream task, leading to improved task-specific performance.

One significant challenge in fine-tuning large language models is the substantial GPU memory required to store gradients and optimizer states during parameter updates. The memory demand increases with the models' parameter counts. 
For example, in our preliminary experiment, we fine-tuned two variants of CodeT5+ \citep{DBLP:conf/emnlp/WangLGB0H23} on the assert generation training dataset \citep{DBLP:conf/icse/WatsonTMBP20} (approximately 5K examples) using identical settings and compared their peak memory consumption:
% \lx{For example, in our preliminary experiment, using 5K training samples from the training set of the assert generation task \citep{DBLP:conf/icse/WatsonTMBP20},  we used two variants of CodeT5+ \citep{DBLP:conf/emnlp/WangLGB0H23} under identical settings
% \todo{Note that memory consumption is also related to dataset sizes; so need to say what task and the dataset sizes} 
The 220-million parameter variant required 12.1GB of GPU memory, while the larger 770-million parameter variant consumed 37.7GB.
Such memory demand challenge is increasingly pronounced as recent models \citep{DBLP:conf/acl/ZhengZSLLFCY24, DBLP:conf/iclr/LuoX0SGHT0LJ24, DBLP:conf/iclr/MuennighoffLZZH24} continue to grow, often containing orders of magnitude more parameters than their predecessors from just a few years ago \citep{DBLP:conf/emnlp/FengGTDFGS0LJZ20, DBLP:conf/iclr/GuoRLFT0ZDSFTDC21, DBLP:conf/naacl/DevlinCLT19}.

% \begin{figure}[h]
%     \centering
%     \includegraphics[width=\linewidth]{img/memory_usage_vs_ft.png}
%     \caption{(a) Comparison of GPU memory requirements to fine-tune CodeT5+ 220M and CodeT5+ 770M using full parameter tuning and \approach (ours). \approach demonstrates significant memory reductions compared to full parameter fine-tuning, where the 770M model achieves up to 87\% reduction (from 37.7 GB to 8.7 GB) while the 220M model shows up to 48\% reduction (from 12.1 GB to 6.3 GB). (b) Comparison with other efficient fine-tuning methods showing \approach achieves the lowest memory usage, requiring 20-42\% less memory than Prefix-Tuning and LoRA across both model sizes.}
%     \label{fig:vs_ft}
% \end{figure}

Prior studies have proposed various efficient adaptation techniques, such as Adapter-based \citep{DBLP:conf/emnlp/BapnaF19, DBLP:conf/icml/HoulsbyGJMLGAG19, DBLP:conf/emnlp/PfeifferVGR20, DBLP:conf/emnlp/PfeifferRPKVRCG20, DBLP:conf/eacl/PfeifferKRCG21, DBLP:conf/nips/LiuTMMHBR22, LoRA, DBLP:conf/iclr/Hyeon-WooYO22, DBLP:conf/iclr/YehHGYO024, DBLP:journals/corr/abs-2202-13914, DBLP:conf/iclr/KopiczkoBA24} and Prompt-based \citep{PrefixTuning, PromptTuning, DBLP:conf/acl/LiuJFTDY022} methods to address the memory requirement challenge.
Adapter-based methods
%\citep{DBLP:conf/emnlp/BapnaF19, DBLP:conf/icml/HoulsbyGJMLGAG19, DBLP:conf/emnlp/PfeifferVGR20, DBLP:conf/emnlp/PfeifferRPKVRCG20, DBLP:conf/eacl/PfeifferKRCG21, DBLP:conf/nips/LiuTMMHBR22, LoRA, DBLP:conf/iclr/Hyeon-WooYO22, DBLP:conf/iclr/YehHGYO024, DBLP:journals/corr/abs-2202-13914, DBLP:conf/iclr/KopiczkoBA24}
introduce additional trainable parameters into a model and update only these parameters during the fine-tuning stage rather than the entire model. On the other hand, embedding-based methods
%\citep{PrefixTuning, PromptTuning, DBLP:conf/acl/LiuJFTDY022}
modify the output of the embedding layer before feeding it to the initial encoder/decoder layer of the model. 

The existing efficient fine-tuning methods
%\citep{DBLP:conf/emnlp/BapnaF19, DBLP:conf/icml/HoulsbyGJMLGAG19, DBLP:conf/emnlp/PfeifferVGR20, DBLP:conf/emnlp/PfeifferRPKVRCG20, DBLP:conf/eacl/PfeifferKRCG21, DBLP:conf/nips/LiuTMMHBR22, LoRA, DBLP:conf/iclr/Hyeon-WooYO22, DBLP:conf/iclr/YehHGYO024, DBLP:journals/corr/abs-2202-13914, DBLP:conf/iclr/KopiczkoBA24, PrefixTuning, PromptTuning, DBLP:conf/acl/LiuJFTDY022}
for code-related tasks in software engineering face two limitations.
First, they involve an inherent trade-off between parameter efficiency and model performance, where reducing trainable parameters can degrade performance compared to full parameter fine-tuning \citep{DBLP:conf/kbse/LiuSP23}. Second, these methods fail to leverage inherent structural and dependency information in source code as they rely solely on the sequential representation of code.
Prior studies have demonstrated that incorporating structural and dependency information can improve model performance compared to learning from code sequences alone \citep{DBLP:conf/icse/ZhangMBBRAJHD23, DBLP:conf/icse/LiuZWL23, DBLP:journals/ese/MiZWBCM23, DBLP:conf/iclr/AllamanisBK18, DBLP:conf/iwpc/ZhangWZ0J22}. Unlike natural language texts, source code contains well-defined structures, dependencies, and control flows that can be explicitly represented using graphs.
These graph-based representations can naturally better capture long-range relationships between program elements than code sequences. For example, connecting variable declarations with their uses or representing control flow relationships that span across many lines.

\
Building on these insights,
% into the limitations of existing fine-tuning methods and the potentials of rich structural and dependency information in code, 
we propose \approach, a new model adaptation and fine-tuning
method for our main goal for code-related tasks: to achieve reasonable performance compared to full parameter fine-tuning by leveraging rich code structural and dependency information using much fewer trainable parameters compared to the other fine-tuning methods.
\approach efficiently adapts large models by leveraging modular neural network layers that minimize trainable parameters while maintaining strong performance.
At the core of \approach is the \transducer, which enriches model inputs using Code Property Graphs (CPGs) \citep{DBLP:conf/sp/YamaguchiGAR14} that capture rich code properties including syntactic structures, control flows, and data dependencies. 
The \transducer comprises two key components: Graph Vectorization Engine (GVE) and Attention-based Fusion Layer (ABFL). GVE extracts CPGs from input source code and converts dependency information into graph feature vectors. ABFL then fuses these vectors with initial code embeddings from an existing model, enriching them with structural and dependency information. \approach optimizes the \transducer for various downstream tasks, improving the input embeddings for the model without fine-tuning the model itself.

We have evaluated \approach on three downstream tasks: code summarization, assert generation, and code translation. The results demonstrate that \approach achieves comparable performance to full fine-tuning while reducing up to 99\% trainable parameters to save GPU memory.
% \todo{better to mention reduced memory costs too}.
Also, \approach delivers competitive results against efficient fine-tuning methods like LoRA \citep{LoRA}, Prompt-Tuning \citep{PromptTuning}, and Prefix-Tuning \citep{PrefixTuning}, while using only 1.5\%-80\% of their trainable parameters.

Our key contributions are as follows:
\begin{itemize}[nosep,leftmargin=1em]
    \item A novel adaptation method, \approach, that effectively adapts large language models for downstream code-related tasks using Code Property Graphs (CPGs) while minimizing trainable parameters and maintaining strong performance. We make the code available at \url{https://github.com/imamnurby/Transducer-Tuning}.
    \item Comprehensive evaluation across three downstream tasks demonstrating that \approach achieves comparable performance to existing methods while significantly reducing the trainable parameters and GPU memory needed.
\end{itemize}

\section{Background on Code Property Graphs}
Code Property Graphs (CPGs) \citep{DBLP:conf/sp/YamaguchiGAR14} unify the Abstract Syntax Tree (AST), Control Flow Graph (CFG), and Program Dependence Graph (PDG). The AST shows the structure of statements and expressions, the CFG outlines the execution order and conditions for code paths, and the PDG captures dependencies between statements, using edges to represent data and control dependencies.

\begin{figure}[t]
    \centering
    \includegraphics[width=0.7\columnwidth]{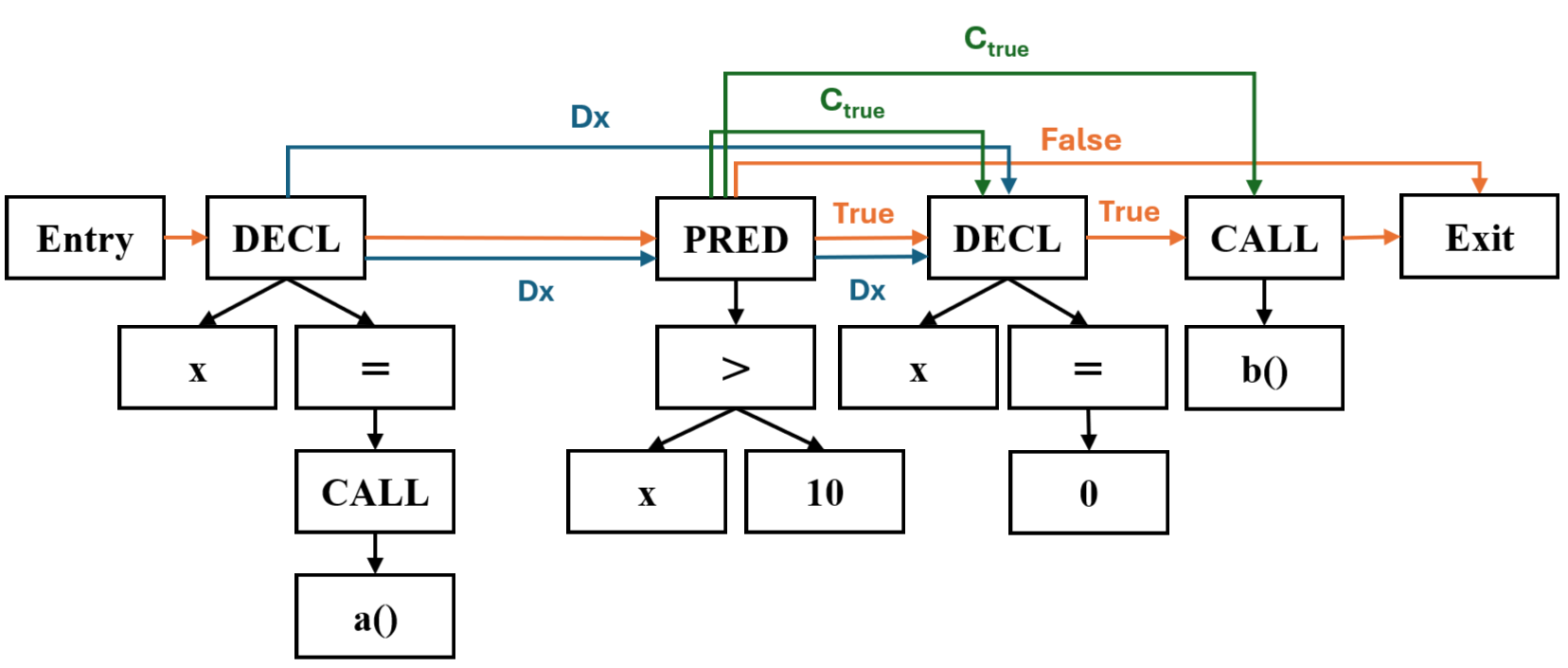}
    \caption{An example of Code Property Graph (CPG). {DECL:} declaration; {PRED:} predicate; {CALL:} function call; {Black Edges:} syntactic relations; {Orange Edges:} control flows; {Blue Edges:} data dependencies; {Green Edges:} control dependencies.}
    \label{fig:cpg-example}
\end{figure}

Figure \ref{fig:cpg-example} illustrates the Code Property Graph (CPG) for the code presented in Listing \ref{lst:code-example}. In this graph, the \texttt{DECL} nodes represent assignments, the \texttt{PRED} node represents the conditional check, and the \texttt{CALL} nodes represent function calls. The black edges correspond to the syntactical structure of the code. The orange edges depict the control flow, illustrating the possible execution paths. Specifically, there are two possible paths: one where the condition \texttt{x > 10} in line 3 is true, and another where it is false. The C$_{\text{true}}$ edges correspond to control dependencies, indicating that the subsequent assignment \texttt{x = 0} in line 4 and function call \texttt{b()} in line 5 depend on the condition \texttt{x > 10} in line 3 being true. The D$_{\text{x}}$ edges represent data dependencies, showing how the value of \texttt{x} is utilized throughout the code.

\begin{listing}[t]%
{\scriptsize
\caption{A simple program for the CPG explanation.}%
\label{lst:code-example}%
\begin{lstlisting}[language=Python]
def main():
    x = a()
    if x > 10:
        x = 0
        b() 
\end{lstlisting}
}
\end{listing}

By jointly taking into account the structure, control flow, dependencies in source code, we believe it can potentially helps language models to achieve a better understanding of source code using fewer trainable parameters during the fine-tuning stage, thus minimizing performance degradation due to using fewer trainable parameters.

\section{\approach}

\subsection{Transducer's Architecture}
\begin{figure}[t]
    \centering
    \includegraphics[width=0.8\columnwidth]{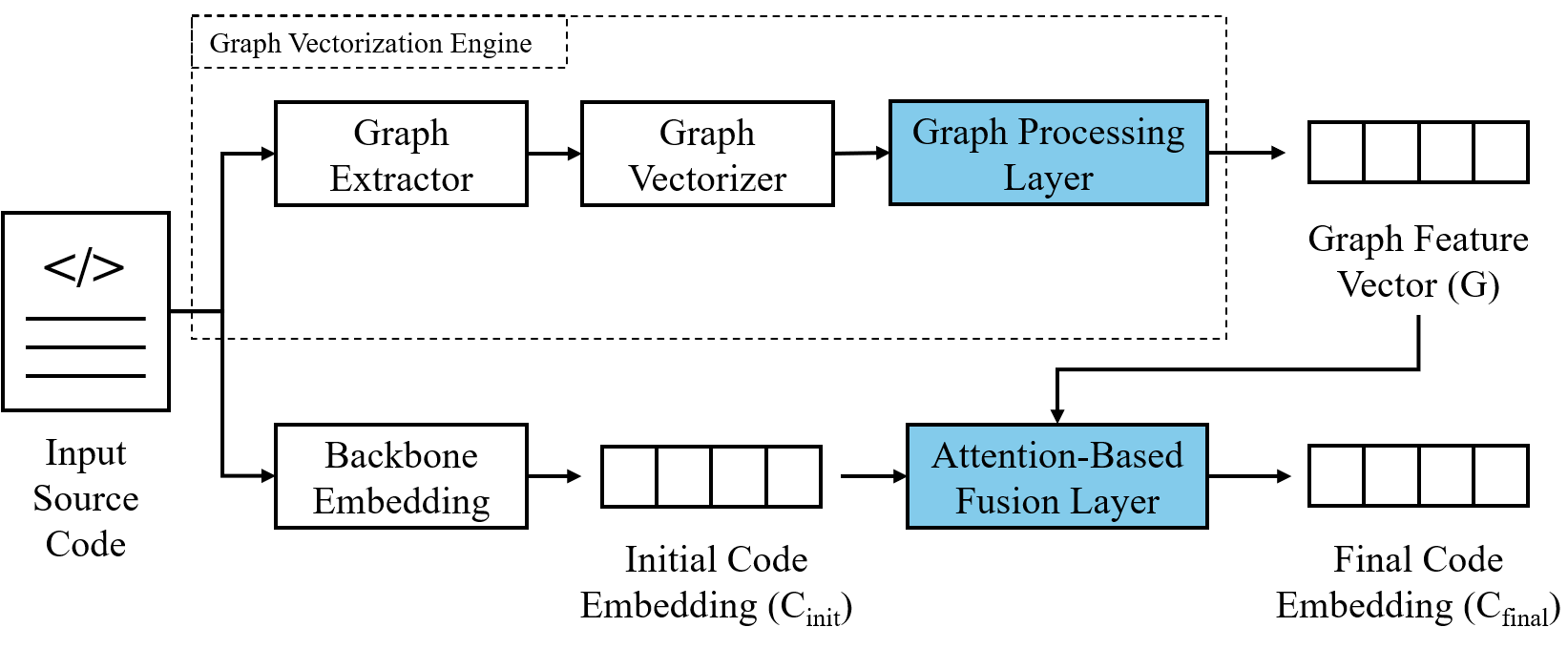}
    \caption{The high-level architecture of \approach, consisting of two main components: (1) Graph Vectorization Engine for transforming source code into graph feature vectors through Graph Extractor, Graph Vectorizer, and Graph Processing Layer, and (2) Attention-Based Fusion Layer for integrating the graph features with code embeddings from a backbone model to produce enriched code embeddings. The blue colored components are updated during fine-tuning.}
    \label{fig:architecture}
\end{figure}

Figure \ref{fig:architecture} shows the high-level architecture of \approach. \transducer  introduces a novel architecture comprising two primary components: Graph Vectorization Engine (GVE) and Attention-Based Fusion Layer (ABFL). GVE processes input source code by extracting and transforming Code Property Graphs (CPGs) into feature vectors. ABFL integrates these features with code embeddings generated by the underlying language model, referred to as the backbone model.

\subsubsection{Graph Vectorization Engine}
\begin{figure}[t]
    \centering
    \includegraphics[width=0.75\linewidth]{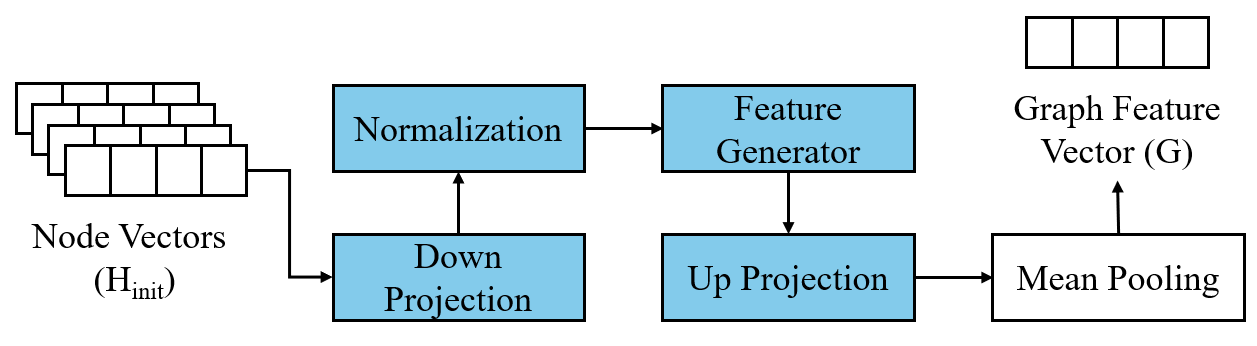}
    \caption{The architecture of the Graph Processing Layer to transform node vectors of CPGs using five components: (1) Normalization for input stabilization, (2) Down Projection for dimensionality reduction, (3) GNN-based Feature Generator for capturing graph relationships, (4) Up Projection for feature reconstruction, and (5) Mean Pooling for generating the final graph representation. The blue colored components are updated during fine-tuning.}
    \label{fig:graph-processing-layer}
\end{figure}

Graph Vectorization Engine (GVE) consists of three subcomponents that work in sequence: Graph Extractor, Graph Vectorizer, and Graph Processing Layer.
(1) The Graph Extractor employs a static code analysis tool to extract the CPG from input code $c$. The CPG is represented as a node list $N$ and edge list $E$.
(2) The Graph Vectorizer then transforms each node label $n_i \in N$ into a vector representation through a mapping function $F: x \rightarrow \mathbf{h}$. The function $F$ can be implemented using various approaches, such as pre-trained embedding models, TF-IDF, or binary vectors. The output is a set of initial node vectors $\mathbf{H}_{\text{init}}$, where each vector $\mathbf{h} \in \mathbf{H}_{\text{init}}$ has dimension $d_\text{init}$.
(3) The Graph Processing Layer transforms the initial vectors into a refined feature representation $\mathbf{G}$ with dimension $d_g$. As illustrated in Figure \ref{fig:graph-processing-layer}, this layer comprises five sequential components: Normalization \citep{DBLP:journals/corr/BaKH16, DBLP:conf/nips/ZhangS19a}, Down Projection, Feature Generator, Up Projection, and Mean Pooling. Each component serves a specific purpose in the processing pipeline.

The first component in the processing pipeline is Normalization, which implements Root Mean Squared (RMS) normalization \citep{DBLP:conf/nips/ZhangS19a}. For each element $h_i$ in the input vector, RMS normalization is computed as:
\begin{equation}
\label{eq:rms}
\mathbf{h}_{\text{norm,i}} = \frac{h_i}{\text{RMS}(\mathbf{h})} g_i, \quad \text{where } \text{RMS}(\mathbf{h}) = \sqrt{\frac{1}{n} \sum_{i=1}^n h_i^2}
\end{equation}
where $n$ is the dimension of the input vector and $g_i$ is a learnable scale parameter. Layer normalization is a technique that stabilizes the distributions of intermediate layer outputs \citep{DBLP:conf/nips/Xu0ZZL19}. By applying Equation (\ref{eq:rms}) to each node vector $\mathbf{h}_{\text{init,i}} \in \mathbf{H}_{\text{init}}$, this layer ensures consistent scaling across all inputs. This normalization has three benefits \citep{DBLP:conf/nips/Xu0ZZL19}: it smooths gradient flow during training, accelerates training convergence, and enhances models' generalization.
%capabilities.

Following Normalization, Down Projection performs dimensionality reduction through a learned transformation. This layer applies trainable weights $\mathbf{W}_{\text{down}}$ to each normalized vector $\mathbf{h}_{\text{norm,i}} \in \mathbf{H}_{\text{norm}}$, producing down-projected vectors $\mathbf{h}_{\text{down,i}} \in \mathbf{H}_{\text{down}}$ with reduced dimension $d_{\text{down}}$, where $d_{\text{down}} < d_{\text{init}}$. This dimensional reduction serves two purposes: it substantially decreases both computational complexity and memory requirements, while simultaneously encouraging the model to learn and retain only the most salient features from the input representation.

Feature Generator processes the compressed representations by transforming the down-projected node vectors $\mathbf{H}_{\text{down}}$ through learnable weights to produce feature vectors $\mathbf{H}_{\text{feature}}$. At its core, this component is implemented as a Graph Neural Network (GNN) \citep{DBLP:journals/nn/JuFGLLQQSSXYYZWLZ24, DBLP:conf/kdd/WuCP0023} that performs iterative message passing between nodes according to the graph structure. During each iteration, nodes exchange information with their neighbors, allowing the model to capture both local and global relationships within a CPG. This message-passing mechanism enables the model to learn rich node representations that reflect not only the node's own features but also its structural context within the graph. Through this process, the transformation extracts and combines relevant features from the reduced-dimensional space, capturing structural and dependency information within the CPG.

Up Projection expands the feature representations by applying learnable weights $\mathbf{W}_{\text{up}}$ to each feature vector $\mathbf{h}_{\text{feature,i}} \in \mathbf{H}_{\text{feature}}$. This generates up-projected vectors $\mathbf{H}_{\text{up}}$ with dimension $d_{\text{up}}$, where $d_{\text{up}} > d_{\text{down}}$. The expansion increases the model's capacity to represent dependencies by projecting the learned features into a higher-dimensional space, where the higher dimension provides the capacity to reconstruct important feature relationships that were compressed during down-projection.

The final graph feature vector $\mathbf{G}$ is obtained through mean pooling, aggregating information across all node vectors in $\mathbf{H}_{\text{up}}$ by computing their element-wise average. This operation transforms the set of individual node representations into a single, fixed-dimensional vector that captures the global characteristics of the entire CPG.
The resulting graph feature vector $\mathbf{G}$ serves as the representation of the input CPG, encapsulating both the local features of individual nodes and their relationships captured during the message passing phase. 
This unified representation is then passed to ABFL to enhance the code embeddings generated by the backbone model.

\subsubsection{Attention-Based Fusion Layer} 

\begin{figure}
    \centering
    \includegraphics[width=0.75\linewidth]{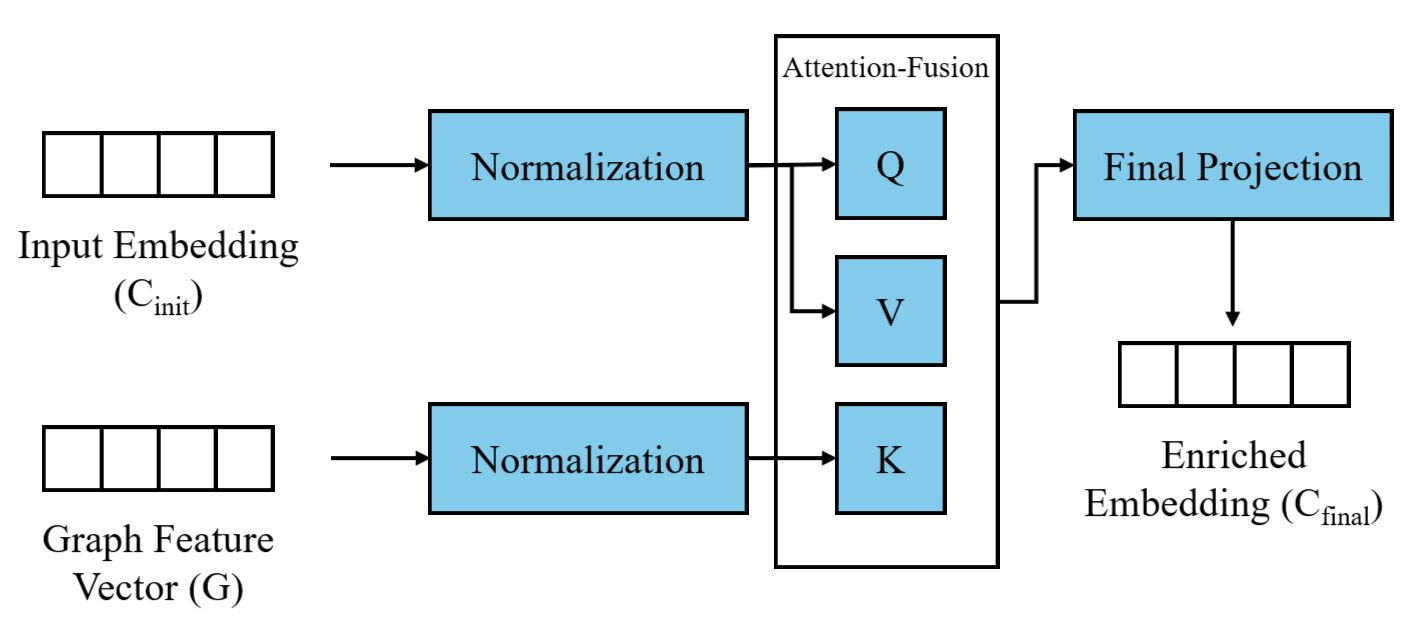}
    \caption{The architecture of the Attention-Based Fusion Layer (ABFL) for integrating graph features with code embeddings using four components: (1) Normalization layers for stabilizing both input embedding and graph feature vector, (2) Attention-Fusion with Query (Q), Key (K), and Value (V) transformations for computing attention weights, and (3) Final Projection for generating the enriched code embedding. The blue colored components are updated during fine-tuning.}
    \label{fig:abfl}
\end{figure}

Figure \ref{fig:abfl} shows the architecture of Attention-Based Fusion Layer (ABFL). ABFL integrates two key inputs: a graph feature vector $\mathbf{G}$ derived from the transducer and an input code embedding $\mathbf{C}_{\text{init}}$ generated by the backbone model's embedding layer. The layer produces an enriched code embedding $\mathbf{C}_{\text{final}}$ that incorporates structural and dependency information encoded in $\mathbf{G}$. ABFL comprises three sublayers: Normalization, Attention-Fusion, and Final Projection.

The process begins with two parallel Normalization that apply RMS normalization according to Equation \ref{eq:rms} to both inputs, producing a normalized graph feature vector $\mathbf{G}_{\text{norm}}$ and a normalized code embedding $\mathbf{C}_{\text{norm}}$. This normalization ensures consistent scaling and stabilizes the distribution of both feature representations. These normalized representations then undergo fusion through Attention Fusion, which implements the attention mechanism introduced in \citep{DBLP:conf/nips/VaswaniSPUJGKP17}.

Attention Fusion uses three trainable weight matrices: $\mathbf{W}_{\text{Q}}$, $\mathbf{W}_{\text{K}}$, and $\mathbf{W}_{\text{V}}$.
%The matrices
$\mathbf{W}_{\text{Q}}$ and $\mathbf{W}_{\text{V}}$ transform $\mathbf{C}_{\text{init}}$ into query vector $\mathbf{Q}$ and value vector $\mathbf{V}$ respectively;
$\mathbf{W}_{\text{K}}$ transforms $\mathbf{G}$ into key vector $\mathbf{K}$. Each resulting vector ($\mathbf{Q}$, $\mathbf{K}$, and $\mathbf{V}$) maintains a dimension of $d_{\text{abf}}$. Then the attention mechanism processes the vectors according to:
\begin{equation}
\label{eq:attention}
\text{Attention}(\mathbf{Q}, \mathbf{K}, \mathbf{V}) = \text{softmax}\left(\frac{\mathbf{Q}\mathbf{K}^\top}{\sqrt{d_{\text{abf}}}}\right)\mathbf{V}
\end{equation}

This computation involves first calculating the scaled dot product between $\mathbf{Q}$ and $\mathbf{K}$, which generates alignment scores determining the importance of each token embedding $\mathbf{c}_{\text{init,i}} \in \mathbf{C}_{\text{init}}$ relative to the graph features in $\mathbf{G}$. These scores are scaled by $\sqrt{d_{\text{abf}}}$ to prevent excessive magnitudes in high-dimensional spaces, then normalized through a softmax function to produce attention weights. The final attention output is computed as a weighted sum of the value vector $\mathbf{V}$ using these normalized weights.

Final Projection concludes the process by transforming the attention mechanism's output into $\mathbf{C}_{\text{final}}$ using weight matrix $\mathbf{W}_{\text{final}}$. Each vector $\mathbf{c}_{\text{final,i}} \in \mathbf{C}_{\text{final}}$ is projected to match the backbone model's hidden dimension $d_{\text{backbone}}$. This enriched code embedding $\mathbf{C}_{\text{final}}$ is then forwarded to the backbone model's encoder/decoder.

\subsection{Usage Scenario}

Using \approach involves two key stages: training and inference. In the training stage, the service provider or end-user selects a backbone model and trains a \transducer component using input-output samples from the target task. During this process, only the \transducer's parameters are updated while the backbone model remains frozen. During inference, the trained \transducer enriches input embeddings that are generated by the backbone model for the target task. When a new downstream task emerges, users can train an additional \transducer while keeping the same backbone model. They simply need to deploy the new \transducer while leaving the backbone model and any existing \transducer unchanged. This modularity also means that any \transducer can be removed when no longer needed without impacting the backbone model or other \transducer that serve different tasks.

\section{Experimental Setting}
\subsection{Datasets}
We evaluate \approach on three downstream tasks: code summarization, assert statement generation, and code translation. We selected datasets that are widely used in recent studies. For code summarization, we use the clean Java subset \citep{clean-CodeSearchNet} from CodeSearchNet \citep{CodeSearchNet}. This dataset contains Java methods with their corresponding Javadoc descriptions. For assert statement generation, we use the dataset created by Watson et al. \citep{DBLP:conf/icse/WatsonTMBP20}. This dataset pairs test methods with their assert statements. For code translation, we use data from CodeXGLUE \citep{DBLP:conf/nips/LuGRHSBCDJTLZSZ21}. This dataset contains parallel code snippets that implement the same functionality in Java and C\#. Each dataset comes from open-source GitHub projects.

\textbf{Dataset Decontamination and Preprocessing.}
We utilize preprocessed and cleaned datasets that have been divided into training, validation, and testing sets by their respective authors. To ensure data integrity, we first check for potential leakage between splits across all datasets. This process involves two stages: first, we remove exact matches between splits, and then we eliminate near-duplicates using Locality Sensitive Hashing (LSH) and MinHash. Specifically, we tokenize each instance and generate a MinHash signature for each one, which efficiently estimates the Jaccard similarity between instances. We then use LSH to group similar items together and remove those with a similarity score greater than 0.8. As a result, 41\% to 53\% of instances are retained from the original test split for code-to-code translation tasks, while 98\% of instances are maintained for code summarization.

Then we extract Code Property Graphs (CPGs) for each method in the datasets using Joern.\footnote{\url{https://github.com/joernio/joern}}. Then, we generate node vectors within the CPGs by converting node labels using the generic embedding model mxbai-embed-large-v1.\footnote{\url{https://huggingface.co/mixedbread-ai/mxbai-embed-large-v1}} We chose mxbai-embed-large-v1 because it performs the best among the small models on the Massive Text Embedding Benchmark \citep{DBLP:conf/eacl/MuennighoffTMR23} at the time of our experiments. Moreover, its small size allows for fast encoding of node labels. 

In the end, the code summarization dataset contains 82K Java methods for training, 9.1K for validation, and 3.1K for testing. The assert statement generation dataset includes 50K instances for training, 6.3K for validation, and 3.3K for testing. The code translation dataset contains 10.3K parallel code snippets for training, 500 instances for validation, and 370 for testing. We release our preprocessed datasets to support reproducibility, with download links provided
in the Availability portion of the appended ``Declarations'' Section.
%in Appendix \ref{app:data-availability}. 
The full dataset statistics are available in Appendix \ref{app:dataset-stats}.

\subsection{\approach Implementation}
We use CodeT5+ \citep{DBLP:conf/emnlp/WangLGB0H23} as the backbone model. We selected this model based on its adoption in recent studies \citep{DBLP:conf/iclr/LuoX0SGHT0LJ24, DBLP:conf/iclr/MuennighoffLZZH24, DBLP:conf/acl/ZhengZSLLFCY24, DBLP:journals/corr/abs-2308-10462} and the availability of smaller variants (i.e., 220M and 770M parameters) that can be fine-tuned on our local workstation. These variants each receive over 10K monthly downloads on HuggingFace as of August 2024. For \approach implementation, we use GATv2 \citep{GATv2} as the Feature Generator, with both the down-projection dimension ($d_{\text{down}}$) and attention-based fusion dimension ($d_{\text{abf}}$) set to 8, which is inspired from the prior study \citep{LoRA}.

\subsection{Baselines} 
We evaluate \approach against both standard and efficient fine-tuning baselines. For standard baselines, we use full fine-tuning (upper bound with all backbone parameters optimized), no fine-tuning (lower bound using only pre-trained state), and Linear Adapter (a linear layer transforming embeddings before the encoder).

We also compare against efficient fine-tuning methods: LoRA \citep{LoRA}, Prefix-Tuning \citep{PrefixTuning}, and Prompt-Tuning \citep{PromptTuning}. For LoRA, we tune the rank (r) of trainable matrices with values {4, 8}, injecting them into the query and key vectors of attention layers. For Prefix-Tuning and Prompt-Tuning, we adjust the prefix length (p) and soft-prompt length (s) respectively, with values {5, 10, 25, 50}. All the values in the search space come from ablation studies in the original papers \citep{LoRA, PrefixTuning, PromptTuning}.

To ensure fair comparison, we tune each method's hyperparameters to achieve optimal performance with minimal trainable parameters. The tuning uses 20\% of the training data and evaluates on the full validation set, with separate tuning for each task and backbone model combination. We report the selected hyperparameters for each task in Appendix \ref{app:hyperparam}. In addition, we detail other hyperparameters and environment settings in Appendix \ref{app:env}.

\subsection{Metrics}
We evaluate \approach against the baselines on both efficiency and performance dimensions. Our primary goal is to minimize the number of trainable parameters while maintaining competitive performance. For efficiency, we compare the number of trainable parameters between \approach and the baselines, following prior work \citep{LoRA, PrefixTuning, PromptTuning}. For performance evaluation, we use the default metrics from the CodeXGLUE benchmark \citep{DBLP:conf/nips/LuGRHSBCDJTLZSZ21}. On code summarization tasks, we measure smoothed BLEU \citep{DBLP:conf/acl/PapineniRWZ02, smoothedBLEU}. For code-to-code translation, we use CodeBLEU \citep{DBLP:journals/corr/abs-2009-10297}. We analyze the relative differences in these scores between \approach and the baselines. To ensure robustness, we run each experiment two times with different random seeds and report the average results.

\section{Results}

\begin{table}[t]
\centering
\begin{tabular}{llccc}
\hline
\multicolumn{1}{c}{\textbf{Model}} & \multicolumn{1}{c}{\textbf{Tuning Method}} & \textbf{Summarization} & \textbf{Assert Generation} & \textbf{Code Translation} \\
\hline
\multirow{7}{*}{\begin{tabular}[c]{@{}p{1cm}@{}}CodeT5+\\\centering 220M\end{tabular}} 
    & Transducer Tuning & 99.84 $\pm$ 0.21 & 82.32 $\pm$ 0.30 & 96.60 $\pm$ 1.31 \\
    & No Finetuning      & 95.49 $\pm$ 0.00 & 76.85 $\pm$ 0.00 & 94.47 $\pm$ 0.00 \\
    & Full Finetuning    & 99.91 $\pm$ 0.01 & 83.16 $\pm$ 0.01 & 97.78 $\pm$ 0.00 \\
    & Linear Adapter     & 98.05 $\pm$ 0.88 & 82.48 $\pm$ 0.02 & 97.70 $\pm$ 0.12 \\
    & LoRA               & 99.91 $\pm$ 0.00 & 83.17 $\pm$ 0.00 & 97.78 $\pm$ 0.00 \\
    & Prefix-Tuning      & 99.93 $\pm$ 0.01 & 83.17 $\pm$ 0.00 & 97.78 $\pm$ 0.00 \\
    & Prompt-Tuning      & 99.91 $\pm$ 0.01 & 83.17 $\pm$ 0.00 & 94.40 $\pm$ 0.27 \\
\hline
\multirow{7}{*}{\begin{tabular}[c]{@{}p{1cm}@{}}CodeT5+\\\centering 770M\end{tabular}} 
    & Transducer Tuning & 98.11 $\pm$ 1.61 & 81.16 $\pm$ 0.71 & 94.88 $\pm$ 0.08 \\
    & No Finetuning      & 87.90 $\pm$ 0.00 & 74.13 $\pm$ 0.00 & 90.10 $\pm$ 0.00 \\
    & Full Finetuning    & 99.81 $\pm$ 0.01 & 83.16 $\pm$ 0.01 & 97.78 $\pm$ 0.00 \\
    & Linear Adapter     & 98.24 $\pm$ 1.39 & 81.23 $\pm$ 0.88 & 97.77 $\pm$ 0.02 \\
    & LoRA               & 99.79 $\pm$ 0.02 & 83.17 $\pm$ 0.00 & 97.78 $\pm$ 0.00 \\
    & Prefix-Tuning      & 99.85 $\pm$ 0.00 & 83.15 $\pm$ 0.02 & 97.78 $\pm$ 0.00 \\
    & Prompt-Tuning      & 99.82 $\pm$ 0.01 & 83.17 $\pm$ 0.00 & 90.15 $\pm$ 0.01 \\
\hline
\end{tabular}
\caption{Performance comparison of different model adaptation methods across code tasks. Results show model BLEU (code summarization) and CodeBLEU (assert generation and code translation), where higher is better with standard deviations across two random seeds. \approach demonstrates substantial improvements over No Fine-tuning baseline while remaining competitive with other tuning methods that require more parameters.}
\label{tab:performance-result}
\end{table}

We evaluate \approach on two task categories: code-to-natural language (code summarization) and code-to-code tasks (assert generation and code translation). Table~\ref{tab:performance-result} presents the performance comparison against the baselines in these tasks.

First, we compare \approach with No-Fine-tuning baseline. For CodeT5+ 220M, \approach improves code summarization by 4.35 points (from 95.49 to 99.84), assert generation by 5.47 points (from 76.85 to 82.32), and code translation by 2.13 points (from 94.47 to 96.60). For CodeT5+ 770M, the improvements are larger: 10.21 points in code summarization,
%(from 87.90 to 98.11),
7.03 points in assert generation,
%(from 74.13 to 81.16),
and 4.78 points in code translation.
%(from 90.10 to 94.88).

\begin{mdframed}[linewidth=1pt]
\textbf{Takeaway 1:} \approach achieves substantial improvements over No-Fine-tuning baselines (2.13-10.21 points) for both CodeT5+ 220M and 770M.
%while maintaining competitive performance against other tuning methods with small gaps (0.07-1.2 points for CodeT5+ 220M)\todo{can remove "while..." as you haven't discussed the gap yet}.
\end{mdframed}
\medskip

Next, we compare performance across tuning methods. For CodeT5+ 220M, \approach (99.84) achieves comparable results to other methods in code summarization, with differences of less than 0.1 points compared to Full Fine-tuning (99.91), LoRA (99.91), Prefix-Tuning (99.93), and Prompt-Tuning (99.91). In assert generation, \approach (82.32) shows slightly lower performance, with gaps ranging from 0.16 to 0.85 points compared to other methods (Linear Adapter: 82.48, Full Fine-tuning: 83.16, LoRA: 83.17, Prefix-Tuning: 83.17, Prompt-Tuning: 83.17). For code translation, while \approach (96.60) lags behind most methods by about 1.1-1.2 points (Full Fine-tuning: 97.78, Linear Adapter: 97.70, LoRA: 97.78, Prefix-Tuning: 97.78), it outperforms Prompt-Tuning (94.40) by 2.2 points.

For CodeT5+ 770M, \approach shows consistent performance with slightly larger gaps. In code summarization, it scores 98.11, approximately 1.7 points below other methods (Full Fine-tuning: 99.81, LoRA: 99.79, Prefix-Tuning: 99.85, Prompt-Tuning: 99.82). In assert generation, it achieves 81.16, about 2.0 points lower than alternatives (Full Fine-tuning: 83.16, LoRA: 83.17, Prefix-Tuning: 83.15, Prompt-Tuning: 83.17). In code translation, while \approach (94.88) trails most methods by 2.9 points (Full Fine-tuning: 97.78, Linear Adapter: 97.77, LoRA: 97.78, Prefix-Tuning: 97.78), it surpasses Prompt-Tuning (90.15) by 4.73 points.

% \begin{mdframed}[linewidth=1pt]
% \textbf{Takeaway 2:} \approach achieves competitive performance against other tuning methods with small gaps (0.1-1.2 points for CodeT5+ 220M and 1.7-2.9 points for CodeT5+ 770M) across code summarization, assert generation, and code translation tasks, while consistently outperforming Prompt-Tuning in code translation by 2.2-4.73 points.
% \end{mdframed}

Notably, the performance gaps between \approach and other tuning methods are substantially smaller than its improvements over No Fine-tuning. For CodeT5+ 220M, while the gaps with other methods are at most 1.2 points, \approach achieves gains of 4.35 points in summarization (99.84 vs 95.49), 5.47 points in assert generation (82.32 vs 76.85), and 2.13 points in code translation (96.60 vs 94.47) compared to No Fine-tuning. The contrast is even more pronounced for CodeT5+ 770M, where despite gaps of up to 2.9 points with other methods, \approach demonstrates remarkable improvements over No Fine-tuning: 10.21 points in summarization (98.11 vs 87.90), 7.03 points in assert generation (81.16 vs 74.13), and 4.78 points in code translation (94.88 vs 90.10). These results indicate that \approach effectively adapts the models while maintaining competitive performance compared to more parameter-intensive tuning methods.

\begin{mdframed}[linewidth=1pt]
\textbf{Takeaway 2:} The performance gaps between \approach and other tuning methods (1.2 points for CodeT5+ 220M and 2.9 points for CodeT5+ 770M) are significantly smaller than its improvements over No Fine-tuning (2.13-5.47 points for CodeT5+ 220M and 4.78-10.21 points for CodeT5+ 770M), demonstrating effective model adaptation with minimal performance trade-off.
\end{mdframed}
\medskip

\begin{table}[t]
\begin{tabular}{clrrr}
\hline
\multicolumn{1}{c}{\textbf{Model}} & \multicolumn{1}{c}{\textbf{Tuning Method}} & \textbf{Summarization} & \textbf{Assert Generation} & \textbf{Code Translation} \\
\hline
\multirow{6}{*}{\begin{tabular}[c]{@{}p{1cm}@{}}CodeT5+\\\centering 220M\end{tabular}} & Transducer Tuning & 30.7K & 30.7K & 30.7K \\
                              & Full Fine-tuning    & 222,882K & 222,882K & 222,882K \\
                              & Linear Adapter      & 589.8K & 589.8K & 589.8K \\
                              & LoRA                & 884.7K & 442.4K & 884.7K \\
                              & Prefix-Tuning       & 184.3K & 921.6K & 184.3K \\
                              & Prompt-Tuning       & 38.4K & 76.8K & 38.4K \\
\hline
\multirow{6}{*}{\begin{tabular}[c]{@{}p{1cm}@{}}CodeT5+\\\centering 770M\end{tabular}} & Transducer Tuning & 37.1K & 37.1K & 37.1K \\
                              & Full Fine-tuning    & 737,639K & 737,639K & 737,639K \\
                              & Linear Adapter      & 1,048.6K & 1,048.6K & 1,048.6K \\
                              & LoRA                & 2,359.3K & 1,179.6K & 2,359.3K \\
                              & Prefix-Tuning       & 491.5K & 491.5K & 491.5K \\
                              & Prompt-Tuning       & 102.4K & 102.4K & 102.4K \\
\hline
\end{tabular}
\caption{Comparison of trainable parameters (K) required by different model adaptation methods. \approach consistently uses the fewest parameters across all tasks, while other methods require significantly larger parameter counts. The variation in parameter counts for LoRA, Prefix-Tuning, and Prompt-Tuning reflects task-specific hyperparameter optimization.}
\label{tab:parameter-result}
\end{table}

Table \ref{tab:parameter-result} demonstrates the parameter efficiency of \approach compared to other tuning methods. For CodeT5+ 220M, \approach requires only 30.7K trainable parameters across all tasks, which is substantially lower than other methods. Full Fine-tuning uses 222,882K parameters, requiring over 7,000 times more parameters than \approach. Linear Adapter needs 589.8K parameters, approximately 19 times more than \approach. LoRA's parameter count varies by task, ranging from 442.4K to 884.7K parameters, representing 14-29 times more parameters than \approach. Prefix-Tuning uses between 184.3K and 921.6K parameters, or 6-30 times more parameters. Even Prompt-Tuning, the most parameter-efficient baseline, requires 38.4K-76.8K parameters, still using 1.25-2.5 times more parameters than \approach.
The efficiency gains are even more pronounced for CodeT5+ 770M. \approach maintains a lean parameter count of 37.1K across all tasks, while Full Fine-tuning requires 737,639K parameters—nearly 20,000 times more parameters. Linear Adapter uses 1,048.6K parameters, about 28 times more than \approach. LoRA's parameter usage ranges from 1,179.6K to 2,359.3K, representing 32-64 times more parameters. Prefix-Tuning consistently uses 491.5K parameters, about 13 times more than \approach, while Prompt-Tuning requires 102.4K parameters, still 2.8 times more than \approach.

\begin{mdframed}[linewidth=1pt]
\textbf{Takeaway 3:} \approach demonstrates superior parameter efficiency, using only 30.7K-37.1K parameters across all tasks and models, while other methods require 1.25-20,000 times more parameters. This efficiency is consistent across different model sizes and maintains competitive performance as shown in previous results.
\end{mdframed}

\section{Discussion}

\subsection{The Usefulness of Graph Information}

To evaluate the impact of incorporating graph information in our approach, we conducted experiments with three variants. The first variant, "GVE + ABFL," represents our complete approach using Graph Vectorization Engine (GVE) with Graph Neural Networks (GNN) and Attention-Based Fusion Layer. The second variant, "GVE-only," removes the ABF layer and combines the graph features from GNN directly with the input code embeddings through summation. The third variant, "ABFL-only," excludes the GVE component to assess the model's performance without graph information.

Table \ref{tab:ablation} shows the results of our ablation study. For CodeT5+ 220M, both GVE+ABF and GVE-only variants outperform the ABFL-only variant in all tasks, with GVE+ABF achieving 96.60 in code translation compared to ABFL-only's 92.53, and 99.84 in code summarization compared to ABFL-only's 94.33. In assert generation, GVE+ABF and GVE-only reach 82.32 and 83.14 respectively, while ABFL-only achieves 77.07.
For CodeT5+ 770M, the impact of graph information varies by task. In code translation, GVE+ABF (94.88) and GVE-only (97.78) outperform ABFL-only (91.11) by 3.77-6.67 points, while in code summarization and assert generation, ABFL-only performs marginally better by 0.53 and 2 points respectively.

\begin{table}[t]
\centering
\begin{tabular}{llccc}
\hline
\textbf{Model} & \multicolumn{1}{c}{\textbf{Variant}} & \textbf{Summarization} & \textbf{Assert Generation} & \textbf{Code Translation} \\
\hline
\multirow{3}{*}{\begin{tabular}[c]{@{}p{1cm}@{}}CodeT5+\\\centering 220M\end{tabular}}  
    & GVE + ABFL      & \textbf{99.84} $\pm$ 0.21 & 82.32 $\pm$ 0.30 & \textbf{96.60} $\pm$ 1.31 \\
    & GVE-only   & 99.31 $\pm$ 0.06 & \textbf{83.14} $\pm$ 0.04 & 96.03 $\pm$ 0.00 \\
    & ABFL-only   & 94.33 $\pm$ 7.32 & 77.07 $\pm$ 8.62 & 92.53 $\pm$ 2.65 \\
    % & Linear   & 99.54 $\pm$ 0.09 & 83.08 $\pm$ 0.07 & 91.76 $\pm$ 0.36 \\
\hline
\multirow{3}{*}{\begin{tabular}[c]{@{}p{1cm}@{}}CodeT5+\\\centering 770M\end{tabular}} 
    & GVE + ABFL      & 98.11 $\pm$ 1.61 & 81.16 $\pm$ 0.71 & 94.88 $\pm$ 0.08 \\
    & GVE-only   & 96.23 $\pm$ 1.55 & 78.79 $\pm$ 2.51 & \textbf{97.78} $\pm$ 0.00 \\
    & ABFL-only   & 98.64 $\pm$ 0.52 & \textbf{83.16} $\pm$ 0.01 & 91.11 $\pm$ 0.33 \\
    % & Linear   & \textbf{99.17} $\pm$ 0.42 & 79.31 $\pm$ 4.64 & 90.40 $\pm$ 0.02 \\
\hline
\end{tabular}
\caption{Ablation study results showing the impact of different components in \approach on model performance across various tasks. The table compares the default setting (GVE + ABFL) with three other variants: GVE-only, ABFL-only, and Linear.
% Results indicate the significance of using information from CPGs and processing it using GNN layer to enhance model performance.
}
\label{tab:ablation}
\end{table}

These results demonstrate the benefits of incorporating structural and dependency information from CPGs. Models using CPG information outperform those without it in 4 out of 6 settings, which includes all three tasks in CodeT5+ 220M and code translation in CodeT5+ 770M. The remaining two settings with CodeT5+ 770M show minimal performance differences. These results indicate that enriching input embeddings with structural and dependency information enhances model performance.

\subsection{Generalizibility of \approach}
While \approach requires input data that can be represented as a graph, it offers broader applicability as \approach is
%graph-agnostic. The approach is 
not limited to Code Property Graphs (CPGs) and can work with various graph structures such as other code graphs, social network graphs, or knowledge graphs. This flexibility means \approach can adapt large language models across diverse domains where relationships between elements can be captured in graph forms. The key requirement is that the input graph data can be represented as adjacency matrix.

\subsection{The Choice of Code Property Graphs}
The choice of Code Property Graphs (CPGs) as our main graph modality is guided by prior studies \citep{DBLP:journals/ase/HanHSLL23, DBLP:journals/corr/abs-2404-14719, DBLP:conf/icse/LiuZWL23} which suggest that CPGs offer richer information compared to other types of code graphs and can improve performance. The primary focus of our work is to explore whether graph modality can be effectively used for efficient model adaptation, rather than determining the optimal graph representation for code. While we acknowledge that an ablation study comparing different types of graphs, such as CST or control-flow enriched CST, could provide valuable insights into the effectiveness of \approach, such comparison is beyond the scope of our current investigation of graph-based efficient fine-tuning. It is worth noting that different graph representations would require different graph extractors (e.g., Joern for CPGs, tree-sitter for ASTs), but these extractors are modular components that can be easily swapped without affecting the core architecture of our approach. We leave this comprehensive exploration of different graph representations for future work.

\subsection{Threats To Validity}
\textbf{Internal Validity.} Hyperparameter selection could affect our study's internal validity. We managed this risk by tuning hyperparameters for each task with a validation split. We also controlled random variation in our experiments. We used a fixed random seed and ran each model configuration twice with different seeds to verify consistency.

\textbf{External Validity.} Our study's external validity faces a risk from data leakage between dataset splits. We addressed this by removing duplicates from the data. We used Locality Sensitive Hashing and MinHash to detect both exact matches and similar instances. The test data might overlap with the pretraining data. Rather than making absolute performance claims, we compared the improvement that \approach achieved over a No Fine-tuning baseline. This relative comparison helps isolate the actual benefits of our method, regardless of any potential data overlap between test and pretraining sets.

\textbf{Construct Validity.} Our choice of metrics could impact the study's construct validity. BLEU and CodeBLEU are standard metrics in code tasks \citep{10.1145/3597503.3639183, DBLP:journals/ese/HuLXLJ20, DBLP:conf/iwpc/YusufJ022, DBLP:conf/msr/YusufJJ23, DBLP:conf/icse/DeyVGD22}. These metrics may miss some aspects of model performance. Yet this limitation does not affect our core findings. Our goal is to show that we can reduce parameters without losing performance. We focus on relative performance differences between models. The absolute BLEU and CodeBLEU scores matter less for this comparison.
 
\section{Related Works}
\label{sec:related-works}

Several techniques have been proposed to adapt pretrained models for downstream tasks, with direct fine-tuning being a common approach. This method updates all model parameters using task-specific data, typically a small set of input-output examples. Direct fine-tuning has proven successful across various software engineering tasks: code repair \citep{DBLP:conf/icse/MastropaoloSCNP21, DBLP:conf/icse/JiangL021, DBLP:conf/kbse/TianLKKLKB20}, code generation \citep{DBLP:conf/iwpc/YusufJ022, DBLP:conf/msr/YusufJJ23}, code mutant injection \citep{DBLP:conf/icse/MastropaoloSCNP21}, code summarization \citep{DBLP:conf/nips/WeiL0FJ19}, assert generation \citep{DBLP:conf/icse/WatsonTMBP20}, and vulnerability detection \citep{DBLP:journals/tse/ChakrabortyKDR22, DBLP:conf/sigsoft/FuTLNP22}. However, as models grow larger, memory requirements increase due to the growing number of trainable parameters. This challenge has led to the development of more efficient adaptation techniques, primarily Adapter-based and Prompt-based methods.

Adapter-based methods \citep{DBLP:conf/emnlp/BapnaF19, DBLP:conf/icml/HoulsbyGJMLGAG19, DBLP:conf/emnlp/PfeifferVGR20, DBLP:conf/emnlp/PfeifferRPKVRCG20, DBLP:conf/eacl/PfeifferKRCG21, DBLP:conf/nips/LiuTMMHBR22, LoRA, DBLP:conf/iclr/Hyeon-WooYO22, DBLP:conf/iclr/YehHGYO024, DBLP:journals/corr/abs-2202-13914, DBLP:conf/iclr/KopiczkoBA24} introduce additional trainable parameters into the backbone model. During adaptation, only these new parameters are updated, leaving the original model unchanged. However, this approach requires careful consideration of parameter placement, demanding knowledge of the model architecture. In contrast, \approach simplifies adaptation by modifying only the input embeddings before processing through the encoder and/or decoder. This makes \approach easily applicable to any existing language model without requiring end-users to understand the model's internal architecture.

Prompt-based methods \citep{PrefixTuning, PromptTuning, DBLP:conf/acl/LiuJFTDY022} append trainable soft-token parameters to the embeddings generated by large models. During training, only these soft-tokens are updated while the model's weights remain frozen. While this approach shares similarities with \approach in modifying input embeddings, \approach is uniquely designed to incorporate graph data (e.g., CPGs) into the embeddings.

\section{Conclusion}
\label{sec:conclusion}
We present \approach, a novel technique for adapting large models to downstream code tasks using Code Property Graphs (CPGs). At its core, \approach uses a modular Transducer component that enriches code embeddings with structural, control-flow, and dependency information extracted from source code. \transducer has two key components: Graph Vectorization Engine (GVE), which converts CPGs into graph feature vectors, and the Attention-Based Fusion Layer (ABFL), which integrates these vectors with initial code embeddings. By optimizing only the \transducer component for each task, \approach enhances model input embeddings without requiring task-specific fine-tuning of the underlying model. Our experimental results demonstrate that \approach achieves comparable performance to full parameter fine-tuning and existing efficient fine-tuning methods while using significantly fewer parameters, making it easier for users to adapt large language models in resource-constrained settings.

In future work, we plan to explore two key directions. First, we will investigate alternative code features beyond CPGs for adapting large language models with \approach, as different features may prove more effective for specific downstream tasks. Second, we will study the transferability of these features across programming languages, with particular emphasis on low-resource scenarios where training data is limited. These investigations could provide valuable insights for improving model adaptation in diverse software engineering tasks.

\bibliography{sn-article}% common bib file

% \pagebreak

\section*{Declarations}
\label{sec:decl}
% *Funding - (information that explains whether and by whom the research was supported)
% *Conflicts of interest/Competing interests - (include appropriate disclosures)
% *Ethics approval - (include appropriate approvals or waivers)
% *Consent to participate - (include appropriate consent statements)
% *Consent for publication - (appropriate statements regarding publishing an individual’s data or image)
% *Availability of data and material - (data transparency)
% *Code availability - (software application or custom code)
% {I copied from another paper. We need to modify this declarations section.}

\textbf{Author Contribution:} The two authors discussed the original idea and refined various details during the project. The first author collected all the data, developed all the code, performed all the experiments, and wrote most parts of the paper. The second author supervised the project and helped to refine the idea, prioritize experiments and revised the writing.

\medskip
\noindent
\textbf{Funding:}
This research is supported by the Ministry of Education (MOE), Singapore under its Academic Research Fund Tier 3 (Award ID: MOET32020-0004) and the scholarship for PhD students from School of Computing and Information Systems (SCIS) at Singapore Management University (SMU). Any opinions, findings and conclusions or recommendations expressed in this material are those of the author(s) and do not reflect the views of Ministry of Education, Singapore.

\medskip
\noindent
\textbf{Conflicts of interest/Competing interests:} The domain of each institution (semicolon separated) that the authors have a conflict of interest with includes \url{smu.edu.sg}.

\medskip
\noindent
\textbf{Ethics approval and Consent to participate:} This article does not require permission for ethics approval or consent to participation as this work is all based on publicly available datasets and does not involve human users or animals in the study.

\medskip
\noindent
\textbf{Consent for publication:} All authors of this manuscript consent to its publication.

\medskip
% \section{Code Availability}
% \label{app:code-availability}
\noindent
\textbf{Availability of code, data and material:}
The source code for our approach, dataset preprocessing, running experiments, and conducting analysis are available at the following URL \url{https://github.com/imamnurby/Transducer-Tuning}. The repository contains the following:

\begin{itemize}
    \item \texttt{analysis} — contains the source code for conducting analysis.
    \item \texttt{preprocessing\_datasets} — contains the source code for dataset preprocessing.
    \item \texttt{src} — contains the implementation of our approach, training script, inference script, and metric computation script.
\end{itemize}

\medskip
\noindent
Our preprocessed datasets can also be downloaded from the following URLs:
\begin{itemize}
    \item Summarization: \url{https://zenodo.org/records/11652923}
    \item Assert Generation: \url{https://zenodo.org/records/11663635} 
    \item Code Translation: \url{https://zenodo.org/records/11664442}
\end{itemize}

\noindent
The raw datasets before preprocessing were adopted from the previous work by Niu et al. \citep{DBLP:conf/icse/NiuLNCGL23}, and can be downloaded from \url{https://github.com/NougatCA/FineTuner}.

\begin{appendices}

% \section{Dataset Availability}
% \label{app:data-availability}
% The raw datasets were adopted from the previous work by Niu et al. \citep{DBLP:conf/icse/NiuLNCGL23}, and can be downloaded from \url{https://github.com/NougatCA/FineTuner}. We preprocessed and decontaminated these raw datasets and have uploaded the final versions to Zenodo. The download links for the datasets are provided below.

\clearpage

\section{Dataset Statistics}
\label{app:dataset-stats}
\begin{table}[h]
\centering
\begin{tabular}{|l|c|c|c|}
\hline
 \multicolumn{1}{|c|}{\textbf{Metric}} & \textbf{Train} & \textbf{Validation} & \textbf{Test} \\ \hline
\textbf{Total} & 82144 & 9147 & 3210 \\ \hline
\textbf{Average \#Node} & 28.441 & 28.344 & 27.36 \\ \hline
\textbf{Average \#Edge} & 71.591 & 71.257 & 67.874 \\ \hline
\textbf{Average \#Token Input} & 67.403 & 67.28 & 67.619 \\ \hline
\textbf{Average \#Token Truth} & 15.395 & 15.522 & 15.366 \\ \hline
\textbf{Max \#Node} & 50 & 50 & 50 \\ \hline
\textbf{Max \#Edge} & 275 & 228 & 193 \\ \hline
\textbf{Max \#Token Input} & 398 & 381 & 387 \\ \hline
\textbf{Max \#Token Truth} & 257 & 134 & 167 \\ \hline
\end{tabular}
\caption{Dataset Statistics for Code Summarization}
\label{tab:code_summarization}
\end{table}

\begin{table}[h]
\centering
\begin{tabular}{|l|c|c|c|}
\hline
\multicolumn{1}{|c|}{\textbf{Metric}} & \textbf{Train} & \textbf{Validation} & \textbf{Test} \\ \hline
\textbf{Total} & 50661 & 6356 & 6262 \\ \hline
\textbf{Average \#Node} & 25.97 & 25.979 & 25.845 \\ \hline
\textbf{Average \#Edge} & 57.979 & 58.067 & 57.676 \\ \hline
\textbf{Average \#Token Input} & 128.633 & 129.067 & 128.142 \\ \hline
\textbf{Average \#Token Truth} & 23.545 & 23.655 & 23.59 \\ \hline
\textbf{Max \#Node} & 50 & 50 & 50 \\ \hline
\textbf{Max \#Edge} & 153 & 148 & 144 \\ \hline
\textbf{Max \#Token Input} & 399 & 398 & 399 \\ \hline
\textbf{Max \#Token Truth} & 482 & 272 & 213 \\ \hline
\end{tabular}
\caption{Dataset Statistics for Assert Generation}
\label{tab:assert_generation}
\end{table}

\begin{table}[h]
\centering
\begin{tabular}{|l|c|c|c|}
\hline
\multicolumn{1}{|c|}{\textbf{Metric}} & \textbf{Train} & \textbf{Validation} & \textbf{Test} \\ \hline
\textbf{Total} & 9176 & 442 & 896 \\ \hline
\textbf{Average \#Node} & 18.421 & 18.436 & 19.188 \\ \hline
\textbf{Average \#Edge} & 43.26 & 43.31 & 45.466 \\ \hline
\textbf{Average \#Token Input} & 36.045 & 36.367 & 36.889 \\ \hline
\textbf{Average \#Token Truth} & 47.935 & 48.088 & 50.425 \\ \hline
\textbf{Max \#Node} & 50 & 50 & 50 \\ \hline
\textbf{Max \#Edge} & 169 & 150 & 157 \\ \hline
\textbf{Max \#Token Input} & 199 & 138 & 114 \\ \hline
\textbf{Max \#Token Truth} & 215 & 140 & 187 \\ \hline
\end{tabular}
\caption{Dataset Statistics for Code Translation}
\label{tab:code_translation}
\end{table}

\clearpage

\section{Final Baseline-Specific Hyperparameters}
\label{app:hyperparam}
\textbf{LoRA.} The rank $r$ of the injected matrices for each task and backbone model is detailed below.
\begin{table}[ht]
\centering
\small
\begin{tabular}{|c|c|c|c|}
\hline
\textbf{Task} & \textbf{Model} & \textbf{Rank ($r$)} \\ \hline
\multirow{2}{*}{Code Summarization} & codet5p-220m & 8 \\ 
 & codet5p-770m & 8 \\ \hline
\multirow{2}{*}{Assert Generation} & codet5p-220m & 4 \\  
 & codet5p-770m & 4 \\ \hline
\multirow{2}{*}{Code Translation} & codet5p-220m & 8 \\  
 & codet5p-770m & 8 \\ \hline
\end{tabular}
\caption{The tuned ranks of injected matrices for each task and model for LoRA.}
\label{tab:lora_tuning}
\end{table}

\textbf{Prefix-Tuning.} We use the following prefix length $p$ for each task and backbone model.
\begin{table}[ht]
\centering
\small
\begin{tabular}{|c|c|c|}
\hline
\textbf{Task} & \textbf{Model} & \textbf{Prefix Length ($p$)} \\ \hline
\multirow{2}{*}{Code Summarization} & codet5p-220m & 10 \\ 
 & codet5p-770m & 10 \\ \hline
\multirow{2}{*}{Assert Generation} & codet5p-220m & 50 \\ 
 & codet5p-770m & 10 \\ \hline
\multirow{2}{*}{Code Translation} & codet5p-220m & 10 \\ 
 & codet5p-770m & 10 \\ \hline
\end{tabular}
\caption{The tuned prefix lengths for each task and model for Prefix-Tuning.}
\label{tab:prefix_tuning}
\end{table}

\textbf{Prompt-Tuning.} We use the following soft-prompts $s$ length for each task and backbone model.
\begin{table}[ht]
\centering
\small
\begin{tabular}{|c|c|c|}
\hline
\textbf{Task} & \textbf{Model} & \textbf{Soft-Prompt Length ($s$)} \\ \hline
\multirow{2}{*}{Code Summarization} & codet5p-220m & 25 \\ 
& codet5p-770m & 50 \\ \hline
\multirow{2}{*}{Assert Generation} & codet5p-220m & 50 \\ 
& codet5p-770m & 50 \\ \hline
\multirow{2}{*}{Code Translation} & codet5p-220m & 25 \\ 
& codet5p-770m & 50 \\ \hline
\end{tabular}
\caption{The tuned soft-prompt lengths for each task and model for Prompt-Tuning.}
\label{tab:prompt_tuning}
\end{table}

\section{Experimental Settings}
\label{app:env}
\subsection{Hardware Setting}
The experiments are run on a machine with operating system Linux Ubuntu Server 20.04.4, GPU NVIDIA GeForce RTX 3090 24GB, and RAM 64 GB.

\subsection{Software Setting}
The CUDA version used is 12.2, and the GPU driver version is 535.183.01. The packages and their respective versions used to run the experiments are listed in the \texttt{requirements.txt} inside  the code appendix.

\subsection{Hyperparameters}
\subsubsection{Training}
\begin{itemize}
    \item Number of epoch: 1
    \item Training batch size: 8
    \item Validation batch size: 32
    \item Learning rate: 0.0003
    \item Maximum grad norm: 1
    \item Gradient accumulation steps: 1
    \item Mixed precision: bf16
    \item Maximum context length: 400 
    \item Optimizer: AdamW \citep{DBLP:conf/iclr/LoshchilovH19}
    \item Learning rate scheduler: Linear
\end{itemize}

\subsubsection{Inference}
\begin{itemize}
    \item Mixed precision: No
    \item Inference batch size: 4
    \item Decoding mechanism: Beam search
    \item Temperature: 1.0
    \item Top-k: 50
    \item Top-p: 1.0
    \item Early stopping: True 
    \item Maximum generated sequence length: Depends on the longest instance in the target test set. See the dataset statistics in Section \ref{app:dataset-stats}.
\end{itemize}

\subsubsection{Random Seed}
We conducted each experiment (involving 2 different models, 7 tuning methods, and 3 datasets) twice. The random seeds used were 8 and 18. The random seeds can be set in our training script located in \texttt{/src/scripts/run-exp-training-inference.sh} inside the attached code appendix, where an there is option to control the random seed.

\end{appendices}

\end{document}